# Detecting Plagiarism based on the Creation Process

Johannes Schneider, Abraham Bernstein, Jan vom Brocke, Kostadin Damevski, and David C. Shepherd

*Abstract*—All methodologies for detecting plagiarism to date have focused on the final digital "outcome", such as a document or source code. Our novel approach takes the creation process into account using logged events collected by special software or by the macro recorders found in most office applications. We look at an author's interaction logs with the software used to create the work. Detection relies on comparing the histograms of multiple logs' command use. A work is classified as plagiarism if its log deviates too much from logs of "honestly created" works or if its log is too similar to another log. The technique supports the detection of plagiarism for digital outcomes that stem from *unique* tasks, such as theses and *equal* tasks such as assignments for which the same problem sets are solved by multiple students. Focusing on the latter case, we evaluate this approach using logs collected by an interactive development environment (IDE) from more than sixty students who completed three programming assignments.

## I. Introduction

Prominent cases of plagiarism like that of the former German secretary of defense Karl Guttenberg, who plagiarized large parts of his Ph.D. thesis, have helped to increase public awareness of this serious problem. Those who enjoy decision power in politics and industry in particular should be elected or appointed at least in part based on their demonstrated integrity. To this end, a systematic eradication of immoral behavior is needed during education, including a reliable way to detect plagiarism. Software is available to assist in identifying plagiarism, but it can often be defeated by simple manipulation techniques, such as substituting words with synonyms (i.e., rogeting), because such software often detects only exact matches of text. We introduce a novel mechanism that supports identification of plagiarized work by capturing the creation process and comparing the works' generation process, rather than comparing only the final products. The creation process is represented by a log comprised of a sequence of events that are collected automatically during the creation of a digital product.

User behavior and sometimes the internal processes of the software used to make the digital product are tracked

J. Schneider is with the Institute of Information Systems, University of Liechtenstein, Liechtenstein.
E-mail: johannes.schneider@uni.li
A. Bernstein is with Department of Informatics, University of Zurich, Switzerland.
E-mail: bernstein@ifi.uzh.ch
J. vom Brocke is with the Institute of Information Systems, University of Liechtenstein, Liechtenstein.
E-mail: jan.vom.brocke@uni.li
K. Damevski is with the Department of Computer Science, Virginia Commonwealth University, Richmond, VA, 23284, U.S.A.
E-mail: damevski@acm.org
D. Shepherd is with ABB Corporate Research, Raleigh, NC, 27606, U.S.A.
E-mail:david.shepherd@us.abb.com

```
Application.Move Left:=0, Top:=0
Selection.TypeText Text:="Hello,"         Hello,
Selection.TypeParagraph
Selection.TypeText Text:="how were you?"   how are YOU?
Selection.MoveLeft Unit:=wdCharacter, Count:=7
Selection.TypeBackspace
Selection.TypeText Text:="a"
Selection.MoveUp Unit:=wdLine, Count:=1
Selection.MoveLeft Unit:=wdCharacter, Count:=7, Extend:=wdExtend
Selection.Font.Bold = wdToggle
Selection.MoveDown Unit:=wdLine, Count:=1
Selection.TypeBackspace
Selection.MoveRight Unit:=wdCharacter, Count:=3, Extend:=wdExtend
Selection.Font.Italic = wdToggle
Selection.MoveRight Unit:=wdCharacter, Count:=2
Selection.MoveRight Unit:=wdCharacter, Count:=4, Extend:=wdExtend
Selection.MoveLeft Unit:=wdCharacter, Count:=1, Extend:=wdExtend
Selection.Font.Grow
Selection.Font.Grow
Selection.Font.Grow
```

Fig. 1. Log created using the macro recorder in Microsoft Word. The outcome is shown on the upper right.

by recording the events involved in the process. Figure 1, which shows a simple example, illustrates that logs typically contain much more information than the final digital product, as the logs contain the entire history of a document's changes in the order in which they occurred. A log is also more machine-friendly to process than a final product is since a log usually consists of a sequence of events in raw text format, whereas a final product could contain multiple fonts and colors and even graphics. The characteristics of logs provide several opportunities for detecting plagiarism. To the best of our knowledge, the only reliable way to avoid detection by our creation-process-based technique requires either detailed knowledge of the inner workings of the creation software or a significant amount of manual work.

The remainder of the article proceeds as follows: Section II reviews related work, while Section III discusses multiple architectures and technical options for log creation, and Section IV explains how they support a variety of detection and cheating strategies. Section V proposes mechanisms based on histograms of a log's events to detect plagiarism automatically. Although automatic detection is the focus of this work, it may not always yield definite results, so Section VI discusses manual inspection. Section VII evaluates the proposed process using programming assignments in software engineering, and the process's strengths and limitations are presented in Section VIII.

## II. Related Work

Software plagiarism has been discussed from the point of view of students and university staff in [11] and [17], which provide a definition of plagiarism and discuss students' awareness of plagiarism. [11] also mentions that more than 80 percent of surveyed stuff check for plagiarism with about 10 percent using dedicated software, while the others rely on manual inspection.

The comprehensive survey [21] discusses plagiarism in general and also three detection methods that focus on texts: document source comparison (such as word stemming or fingerprinting), search of characteristic phrases and stylometry (exploiting unique writing styles). A taxonomy for plagiarism that focuses on linguistic patterns is given in [3], the authors of which conclude that current approaches are targeted to determine "copy-and-paste" behavior but fail to detect plagiarism that simply presents stolen ideas using different words. Although [21] mentions some electronic tools for detection, surveys that cover these tools and systems (e.g., [20], [5], [23], [25]), reveal techniques that involve computation of similarity for documents (i.e., attribute-counting or cosine distance) [16]. The relationship between paraphrasing and plagiarism has been examined as well (e.g., [23], [38]) as has citation-based plagiarism (e.g., [15], [24], [23]), an approach that uses the proximity and order of citations to identify plagiarism. Other techniques cover parse tree comparison for source code and string tiling [35]. More recent work has identified programmers based on abstract syntax trees with a surprisingly high success rate despite obfuscation of code [8]. However, none of these tools and methods takes the creation process into account.

Several special techniques address source code plagiarism. For example, [4] introduces plagiarism detection of source code by using execution traces of the final program, that is, method calls and key variables. An API-based control flow graph used in [9] to detect plagiarism represents a program as nodes (statements), with an edge between two nodes if there is a transition in the graph between the nodes. API-based control flow graphs merge several statements to an API call, which extracts features and then builds a classifier to detect plagiarism. Focusing on detecting plagiarism for algorithms, [37] relies on the fact that some runtime values are necessary for all implementations of an algorithm, so any modification of the algorithm will also contain these variables. Detection of plagiarism for programming assignments that state similar metrics of source code is investigated in [28], which also provides an extensive experience report. Similarity in [28] is based on counts and sequences of reserved words that assignments have in common.

A comprehensive survey of statistical methods with which to detect cheating is given in [7]. In a multiple choice test, a student takes more time to answer a difficult question than she does to answer a simple question. Statistical response-time methods [7] exploit this effect, which is difficult for cheaters to capture. (We believe that this difficulty also holds for programming and writing a thesis.) Methods that address the intrinsic aspects of a task, such as using latent semantic indexing for text documents [2] and source code [13], have also been proposed that might also be applicable for log analysis. More generally, existing insights on the knowledge-generation process, such as those in [27] and findings for specific tasks and practitioners like novice researchers' writing process [31], could be used in our work as well. Our proposed approach is more inductive (i.e., driven by data) than deductive, as it relies on general insights related to cognitive processes in knowledge creation.

Human-computer interaction can be logged using any of many tools. Our evaluation focuses on logging events from an interactive development environment (IDE), which has a variety of logging tools (e.g., [36], [32], [18], [33]). We decided in favor of Fluorite [36], since it gives fine-grained logging of events. Other tools (e.g., [33]), provide less fine-grained logging with better privacy guarantees and gamification approaches, which are preferable for anonymous usage data collection and evaluation [14]. Mouse movements and events outside the IDE are not logged, and some tools can replay sequences of events [32].

Papers like [34] and [1] use logs to analyze students' learning process. For example, [34] looks at novice students' interactions with a complex IDE to determine how certain characteristics, such as compilation errors, evolve with experience. Students in need of help are identified in [1], who employ a variety of features in machine-learning. The only feature of the code snippets they use is the number of steps taken to complete an assignment. For instance, they compute what compilation errors occurred and how these errors' distribution changes as the students' experience widens. They identify states during the evolution of the source code and correlate student performance with overall course performance. Other work, such as that in [6], identifies patterns using the frequency and size of code updates and performs in-depth analysis of an assignment using a simple language for robot programming.

The study in [30] shows that users can be identified based on the usage of Linux shell commands, techniques that could be helpful in detecting copying of partial logs. The timing between keystrokes [10] can serve as biometric data to identify users, which could be valuable in our context as well. Authorship can also be traced by extracting features like word richness and punctuation from text [19]. The work in [26] provides an evaluation framework for plagiarism detection that focuses on text that, for example, has been changed by random inserts of text and by changes of semantic words. We also use random variations in our creation of artificial logs.

## III. DEFINITIONS AND SCOPE

*Plagiarism* is commonly defined as the "unjust appropriation," "theft and publication" of another person's creation, including "language, thoughts, ideas, or expressions" and the representation of it as one's own. Plagiarism so defined could refer to copying paragraphs from the Internet without citation, slightly modifying parts of a fellow student's for an assignment, paying a third party to do the work, and so on. We focus on the case in which a supposed creator copies significant parts of his or her digital outcome from another source. This "creator" (as a plagiarist is not) might copy literally or create a modified version of the original by means of, for example, word substitution or even by changing the content in a more complex way, such through document structure or semantics, without attribution. This scenario covers the most appealing approach for a plagiarist – copying as much as possible with little modification. It also includes more subtle approaches, such as rewriting an outcome to avoid the time-consuming process of deriving a solution for a given task. We do not attempt to detect when a "creator" copies small portions of a work unless that portion contributes significantly to the overall effort. For instance, we address the scenario in which a student copies most of a course assignment or a thesis and do not attempt to detect when, for example, a short paragraph is copied unless it reflects a large amount of work (e.g., the most difficult part of a programming assignment) and its use manifests in significantly less time and effort (and, thus, in a shorter or different log), in which case we are likely to detect it.

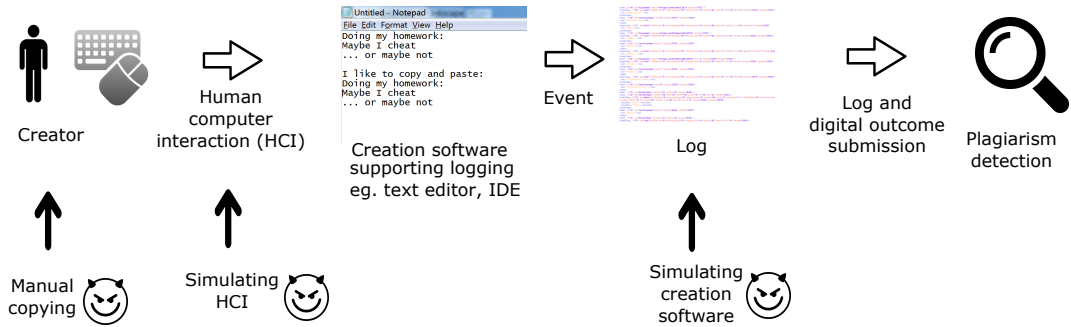

Fig. 2. Log-creation process and three ways for cheating

A *digital outcome* for an assignment or a thesis consists of the files that contain the final product, such as the formatted text of a thesis or source code. The digital outcome can be made by a *creation software* like a programming IDE or a text-processing software and is accompanied by a *log* file that documents the creation process by means of a sequence of automatically logged events. A log often allows the digital outcome to be reconstructed; that is, the events in the log describe the creation process step by step and they resemble all manual editing. In this case, "replaying" the log yields the digital outcome, so it is not strictly necessary to submit a log as well as the digital outcome. We assume both are submitted, since having both aids plagiarism detection by making it possible to compare the digital outcome that stems from replaying the event log and the submitted digital outcome. The creation process can be captured either by the creation software's emitting events or by a general-purpose human-computer interaction (HCI) recorder that tracks mouse and keystrokes. These options are shown in Figures 2 and 3, respectively.

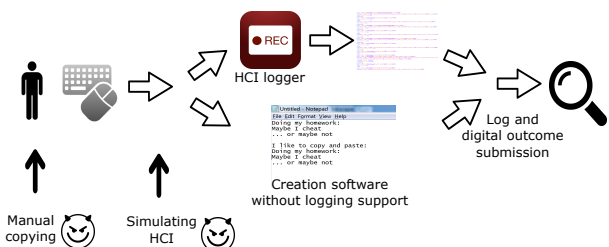

Fig. 3. Log creation by tracking human computer interaction directly

The setup in Figure 3 for tracking HCI is more general and more widely applicable than is reliance on logs from the creation software since not all applications support logging. However, many widely used tools, such as Microsoft Office and OpenOffice applications, come with a recorder for macros (i.e., logs), or they allow installation of plugins to record user activity (e.g., programming IDEs like Eclipse and Visual Studio). Generally, logs that stem from the creation software contain HCI that is categorized into specific events. For instance, a mouse click at a certain position might become a command execution event if the click was on a menu item. Additional events that stem from internal processes might also be logged. Both setups – tracking HCI (in Figure 2) and logging done by the creation software (Figure 3) – allow for similar detection techniques. Tracking HCI raises privacy concerns, particularly if interactions with applications other than the creation software are also logged. Therefore, we focus on the setup shown in Figure 2 and refer to a log as events that stem from the creation software.

```
<DocumentChange _id="368" _type="Insert" docASTNodeCount="108" docActiveCodeLength="658" docExpressionCount="53" docLength="1478" length="34" offset="725" repeat="33">
  <text><![CDATA[servSock = new ServerSocket(p|x|);]]></text>
</DocumentChange>
<Command _id="369" _type="InsertStringCommand" repeat="34" timestamp="19842223" timestamp2="19850601">
  <data><![CDATA[servSock = new ServerSocket(p|x|);]]></data>
</Command>
<Command _id="435" _type="EclipseCommand" commandID=
  "eventLogger.styledTextCommand.COLUMN_PREVIOUS" repeat="4" timestamp="19851010" timestamp2="19851456" />
<DocumentChange _id="439" _type="Delete" docASTNodeCount="108" docActiveCodeLength="657" docExpressionCount="53" docLength="1477" endLine="24" length="1" offset="754" startLine="24" timestamp="19851953">
  <text><![CDATA[i]]></text>
</DocumentChange>
<Command _id="440" _type="EclipseCommand" commandID=
  "eventLogger.styledTextCommand.DELETE_PREVIOUS" timestamp="19851968" />
<DocumentChange _id="441" _type="Insert" docASTNodeCount="108" docActiveCodeLength="658" docExpressionCount="53" docLength="1478" length="1" offset="754" timestamp="19852302">
  <text><![CDATA[0]]></text>
</DocumentChange>
<Command _id="442" _type="InsertStringCommand" timestamp="19852309">
  <data><![CDATA[0]]></data>
</Command>
```

Fig. 4. Log created using a plugin in the Eclipse IDE

An *event* can be triggered immediately either by a human's interacting with the creation software (e.g., by executing a command) or by internal processes of the creation software. Commonly used commands are related to navigation, editing text (insert, replace, delete), selecting menu entries (opening or saving files), and so on. Exemplary events that the creation software generates include "autosave" events (logging that a file was saved automatically), spell-check completed (indicating that a background process for checking the words in the document has finished), and breakpoint hit events in a programming IDE (saying that a program suspended execution at a breakpoint). An event often consists of a timestamp, the name of the event (typically the command type), and details about the event, such as its parameters. For instance, "2015-12-08 10:00.00, InsertString, 'A', position 100:20" corresponds to a user's pressing the "A" key while editing text at line 100, column 20. The exact content of events and the kind of logged events can vary depending on the creation software. Figures 1 and 4 show two examples of log files that originate from different applications.

We use the following *notation* throughout this work. The set of all logs is denoted by $\mathbb{L}$. A log $L \in \mathbb{L}$ consists of a sequence of entries. The $i$-th entry $l_i$ of log $L$, which corresponds to a single logged command or event, is a tuple that consists of a timestamp $t_i$ and an event or command $c_i$ of a certain type. $N(L, c)$ denotes the number of occurrences of command type $c$ in log $L$. (We focus on commands triggered by a human rather than system generated events.) The set of all command types is given by $T$. The types in a single log $L$ are given by $T(L)$. Let $L_U \subset L$ be the subsequence of $L$ that contains only commands of type from the set $U \subseteq T$.

## IV. Cheating and Detection

We elaborate on basic detection mechanisms before defining the challenges that they present to a cheater. We also show differences in the detection of plagiarism in tasks like writing a unique thesis and doing the same assignment as a set of students.

### A. Detection Overview

Detection builds heavily on two mechanisms. The first relies on investigating whether the distribution of events originates from honest creation or plagiarism. The second checks the validity of the log, that is, that the creation software can have produced the log. Ideally, the log can be "replayed" to yield the submitted digital outcome.

The key assumption is that a plagiarized work results in a different sequence and distribution of events than an original work would have, such as fewer edit events and more copy-and-paste events. Our proposed detection system relies on two dimensions: checking for frequencies of a command and checking for command types used. Relying on frequency counts of event types that are extracted from each log, the histogram-based technique computes distances between logs for multiple samples of event types. Thus, we verify: (i) whether a cheater and a non-cheater have event types in common or the cheater uses uncommon command types; and (ii) whether the command types are used approximately the correct number of times–not too similar to any other log or too highly dissimilar from most other logs. In data mining terminology, we perform a nearest-neighbor search to identify manipulated logs, such as those obtained by a plagiarist who changes an honestly created log. We also perform outlier detection to find newly created logs that stem from plagiarized work.

Invalid logs are classified as those that originate from attempted manipulation like plagiarism. For a log to be valid, it must fulfill semantic and syntactic correctness. Syntactic correctness refers to obeying certain rules that determine how events are represented in the log. We say that a log is semantically correct if the creation software could have created it. For example, in an empty text editor, an event for deletion of a character "A" is likely to yield a semantically incorrect log; since the text editor is empty, the creation software could not have emitted such an event. Generally, a cheater must know the state of the creation software, which determines both feasible events and their impact. For example, pressing the "return" key might move the cursor to a new line, but if the software displays a dialog with focus on the "cancel" button, such a key press might close the dialog. It might be possible to verify a log by "replaying" its events, as invalid logs are likely to result in an error message while they are replaying. For instance, standard office tools come with a recorder for macros that can function as logs that can be replayed.

### B. Creating Forged Logs

We discuss the feasibility of and effort required to plagiarize depending on the system architectures and the availability of honestly created logs that could be used to plagiarize. We begin by elaborating on all three ways to cheat (Figure 2). Other ways to cheat, such as manipulating the creation software or the plagiarism detection software itself, seem significantly more challenging, so they are not discussed.

i) Manual Copying: The easiest way to cheat is to interact manually with the creation software. For instance, a student might create a thesis by retyping large parts of text from a website with some rephrasing or by copying and pasting. This approach is easy, and it always yields a semantically and syntactically correct log, but it might come with significant workload for the cheater and/or substantial risk of detection if done carelessly even if the digital outcome itself shows no signs of plagiarism. The distribution of events in the log might deviate from logs of honestly created work, thereby disguising the cheater. Therefore, a successful cheater must anticipate the distribution of events that is typical for honestly created digital outcomes and must forge a log that is close to such a distribution. Essentially, a cheater must perform roughly the same amount of interaction with the creation software as a non-cheater and must interact in a similar way as non-cheaters. For example, for a programming assignment, the cheater must perform a similar degree of program testing and "debugging," which would not be needed if he just wanted to change the structure of the code. If the cheater is writing a thesis, he must perform a similar degree of navigation, redoing, undoing, and editing.

ii) Simulating HCI: A log can also be created indirectly by simulating HCI with the creation software. This approach requires special software that can emit mouse movements and key-press events. The goal for the cheater is to use a partially automatic process to create a log that does not appear to be forged, rather than performing all interactions manually. For example, a student might copy a large part of text from another source and paste it into the creation software, resulting in a single "paste" event in the log that might appear suspicious. On the other hand, he might use a special tool that takes the copied text as input and simulates HCI by sending key-press and mouse-movement events to the creation software, resulting in a more credible log. Automating HCI is possible and an essential part of GUI testing [22] and automation (as showcased by the macro functionality in office applications). Recording the screen coordinates of mouse movements and clicks is sensitive to screen resolutions and software configuration, which reduces portability. For now, even using an available recording and replay tool would require that the cheater configures the tool in order to, for example, replace a single paste command with single key presses of the letters of the text or specify how navigation and editing events should be created by such a tool. A key challenge for the cheater is to be aware of the state of the creation software and the events that are feasible in that state, as well as their effect. For instance, a "cut text" operation is available only when text has been selected. Non-availability of an operation is indicated by a gray "cut" icon and menu item in the text editor's GUI, and clicking on such an icon or menu item has no effect, so the creation software does not generate an event that is appended to the log. Thus, a cheater who attempts to simulate "cut" operations must ensure that the command is available if he is to achieve his goal of appending to the log. In a programming IDE, the user interfaces that result from a click on "Start/Run program" differ depending on whether the program compiles and is executable. If it is not executable, a dialog with an error message might be shown. If it is executable the program might run in the foreground while a toolbar with several debug commands (e.g., for pausing the program to investigate its internal state) appears. If a cheater wants to simulate debugging behavior, he must ensure that the program is executable.

Although simulating HCI always yields a semantically and syntactically correct log, replaying the log might yield a digital product that differs from the submitted digital outcome, identifying the cheater. A human can determine the availability of a command easily by looking at the GUI, but doing so is not so easy for tools, which must rely on keeping track of the internal state of the creation software. Although the cheater might be able to address these challenges, they make forging a log automatically complex.

iii) Simulating the creation software: A cheater might also create or alter the creation software's log directly if he has access to it. In the simplest case, the log is a human-readable text file that allows straightforward manipulation. A cheater might add events or change events in the log to mimic sequences of events that the creation software would emit for honestly created logs. Analogous to simulating HCI, the cheater wishes to manipulate the log file automatically rather than to edit it by hand. To the best of our knowledge, there are no tools with which to create or modify such logs automatically. The challenges are analogous to those for simulating HCI; that is, the cheater must anticipate the software's internal state so the log yields the final digital outcome when it is replayed. What's more, a manipulated log might not even be semantically or syntactically correct, resulting in an error message when it is replayed and immediately revealing a forged log.

Various types of system *architectures* determine whether logs are collected online or offline and whether they are created by software that is running on a device under the creator's control. The architecture determines the ease with which logs can be changed. In the least secure setup, logs are created offline on a student's laptop. If logs are not encrypted and in text format, the student can alter them before submission. In such a setup, a student might use all three forms of cheating. In the most secure technical configuration, a student uses the creation software on a device on which he can neither access logs nor install software that allows HCI to be simulated.

The availability of logs or recordings of honestly created works' HCI facilitates two kinds of forging:

i) Copy and Modify, where a cheater has access to at least one honestly created log that he manipulates before he submits it. To minimize effort, he might take a fellow student's log and perform only a small number of changes. From the perspective of detection this method results in plagiarized logs that are similar to the original log.

ii) Generation from scratch: The cheater does not have access to a log but does have access to at least parts of the digital outcome, such as source code from a fellow student or the Internet, that he plagiarizes by incorporating it in his own work. This method is more likely to result in logs that are dissimilar from all other logs.

### C. Unique vs. Equal Tasks

Our detection method relies on comparing multiple logs. If many students do identical assignments, such as to solve a mathematical problem, we expect similar logs with some variations. Given a unique task, such as writing a thesis, we must rely on comparing the logs of multiple students, each of which writes about a different topic. Detection capability relies on the assumption that writing a thesis conforms to a different process than plagiarizing a thesis does, and that that difference is reflected in the logs through, for example, less editing. Detection is likely to be better if the theses are similar in terms of expected effort and are written by students with similar levels of education.

## V. AUTOMATIC DETECTION BASED ON HISTOGRAMS

We compute frequency statistics for each log and compare them with each other using the counts of each event type, which yield a histogram. If two logs have very similar statistics or a log shows use of commands that differs markedly from those of the other logs, the log is likely to belong to a cheater. The case of two logs with very similar statistics is likely a result of copying a log and modifying it, while the case of markedly different use of commands is likely the result of creating a log from scratch while copying parts of the final digital outcome. The metrics we employ to detect pairs of similar logs and outlier logs use some general design considerations. The detection should be as robust as possible to a cheater's potential modifications; in particular, the cheater's changing a single command type should not impact the metric too much since manually increasing the counts of some commands (e.g., moving the cursor left and right to increase the counts of these two commands) is relatively easy to do. For outlier, low use of commands that are commonly found in honestly created logs are a strong indicator of cheating.

A discussion of general aspects of data preparation is followed by metrics for plagiarism detection.

### A. Data preparation

Data preparation might involve data cleaning and transformation. Data cleaning encompasses removing entire logs as well as cleaning the content of individual logs. Assignments and theses are commonly only partially finished when they are handed in. The corresponding logs might be removed since a cheater is not likely to copy an incomplete assignment. In addition, they are often characterized by their shorter-than-normal length. As a result, they could limit the ability to identify outliers since they do not resemble a completed assignment. For our assignments, we provided a source code skeleton for the students. A few students made minor changes to the skeleton that were far from any serious attempt to complete the task, so the logs contained a sizable portion of navigation but relatively little editing or debugging. A cheater who creates a log with similar properties might escape detection if these logs remain in the dataset.

The content of individual logs typically requires no cleaning since it is generated automatically but data transformation could involve renaming events. For example, we shortened several long command names and ignored all information like event timestamps and parameters, focusing instead only on the command type. We computed a histogram for each log that captures the frequency of each command type. In our data, the distribution of the number of uses $N(L, c)$ of a command $c$ across logs $L \in \mathbb{L}$ is skewed. We transformed the data using the Box-Cox transform to get a more symmetric distribution and added 1 to all values before transformation to handle the case of zero counts, as otherwise the Box-Cox transform might compute $\log 0$, which is undefined. By $B(L, c)$ we denote the transformed value of $1 + N(L, c)$. Figure 10 shows the transformed distribution for frequent command types, while less commonly used command types have a dominating peak at 0.

### B. "Copy and Modify" Detection

The goal of "Copy and Modify" detection is to determine if one log is an altered version of another by computing the similarity between two logs–that is, the Pearson product-moment correlation coefficient of two logs. Using all command types for computation of the correlation might not be robust to insertion or deletion of a few event types, as in the most extreme case, a cheater might alter the frequency of a single command type to such an extent that an otherwise equivalent log is not seen as similar. Therefore, we compute the correlation for randomly chosen subsets of commands. For a pair of logs to be similar it suffices that the correlation for just one of the subsets is similar. There is a risk that if we choose subsets that are too small–too few event types or strongly dependent event types–the distance between two logs will be small by mere chance, but this is a minor concern since only a few commands occur in most of the logs. Therefore, command types must be chosen for correlation computation separately for each pair of logs, rather than choosing just one subset of command types for all logs. Otherwise, if one chooses a subset of rare commands that occurs in only a few logs, many logs will be classified as similar since many logs might not contain any of the rare commands at all. Therefore, to compute the similarity of a pair of logs we select only those commands that occur in at least one of the two logs. However, choosing the subset from only one of the logs ignores the number of different command types that occur in the other log, which might lead to logs' being judged as similar even though one of them contains a significant number of additional event types. Therefore, we choose half of the subset of command types from each log.

More formally, a subset $S$ of commands is chosen as follows for a pair $L, L'$ of logs: We choose half (i.e., $|S|/2 = s_{sam.}/2$) of all command types $U \subset S$ uniformly at random from $L$ (i.e., $T(L)$), and the other half $U' \subset S$ from $T(L') \setminus U$. The entire subset $S$ of command types contained in the histogram is given by $S = U \cup U'$. The Pearson-product correlation $\rho(L, L', S)$ for a subset $S \subset T(L) \cup T(L')$ of command types is given by:

$$\overline{n(L)} := \frac{\sum_{c \in S} n(L, c)}{|S|}$$

$$\rho(L, L', S) := \frac{\sum_{c \in S}(n(L, c) - \overline{n(L)}) \cdot (n(L', c) - \overline{n(L')})}{\sqrt{\sum_{c \in S}(n(L, c) - \overline{n(L)})^2} \cdot \sqrt{\sum_{c \in S}(n(L', c) - \overline{n(L')})^2}}$$

The choice of a subset and the similarity computation is illustrated in Figure 5. In the figure, we chose two commands of each log. The correlation was computed using just four values per log (i.e., one for each of the chosen command types).

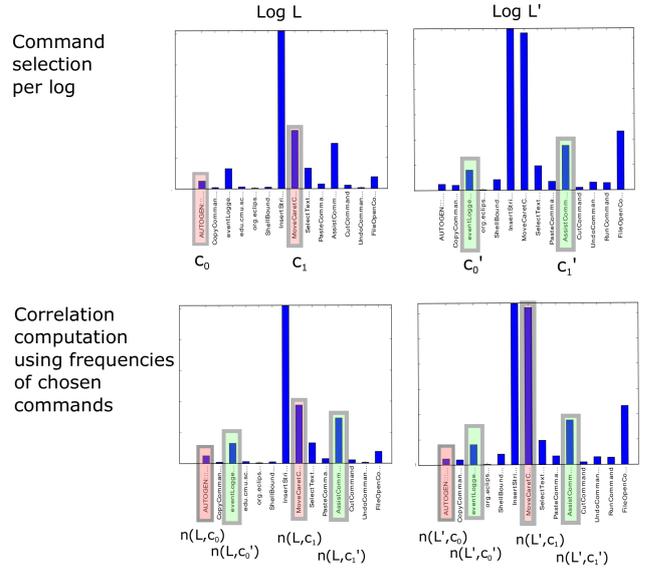

Fig. 5. Illustration of a similarity computation of two logs for "Copy and Modify" detection using one subset

Let $\mathbb{S}(L, L')$ be the set of all chosen subsets $S$ for a pair of logs $L, L'$. The correlation $cor(L, L')$ of two logs is given by the maximal Pearson-product correlation $\rho(L, L', S)$ of any set $S \in \mathbb{S}(L, L')$:

$$cor(L, L') := \max_{S \in \mathbb{S}(L, L')} \rho(L, L', S) \tag{1}$$

Potential cheating candidates are indicated by a larger correlation. For example, we might report a number of pairs of logs having largest correlation among all pairs of logs from $\mathbb{L}$ as cheating candidates.

## C. "Generation from Scratch" Detection

A cheater might not use an existing log for plagiarism but copy a final outcome (without the log). He might also enter a modified solution manually or copy-and-paste it before editing it. In both cases, he creates a log for his plagiarized work. Clearly, given enough effort and time, any final outcome, such as a thesis, can be changed to another outcome whose structure appears to be dissimilar to the original. However, the process of changing such an outcome typically differs from that of solving the task honestly. For example, plagiarizing might be characterized by activities like word substitutions and reordering, changing the work's layout, and (in software engineering) permuting commands. We might expect less incremental changes, rework, navigation, and (in software development) debugging and navigating between files in a plagiarized outcome. Therefore, a log's containing events of some command types that are much more (or less) frequent than they are in most other logs might indicate plagiarism.

Our measure, the outlier score, is a weighted sum of scores of individual commands. The weights are higher for commands that occur in many logs at least once. The score for the "Paste Command" is illustrated in Figure 6, together with the Box-Cox transformed frequency distribution of the number of uses of that command per log. If the frequency of the command's use is normal–that is, within one standard deviation from the mean, the outlier score is zero, but the score rises rapidly to almost one when the frequency of the command's use moves another two standard deviations away from the mean.

Next, we explain the outlier score in more detail and then explain the reasons for choosing each step in the computation.

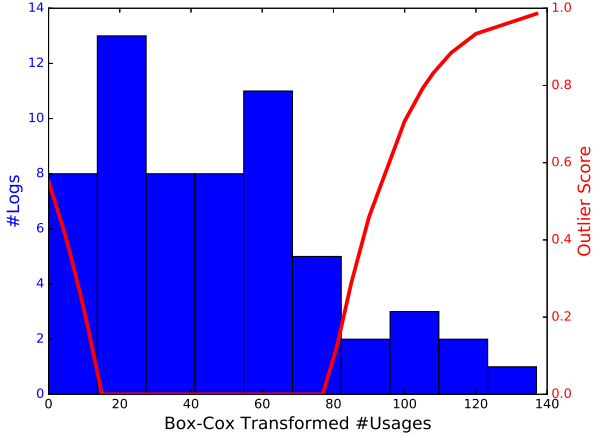

Fig. 6. Contribution of a single command to the outlier score

Roughly speaking, the count of a single command of some log $N(c, L)$ is an outlier if it is much larger or much smaller than the count in most other logs. To get a quantitative estimate, we used the distribution of the Box-Cox transformed counts given by $B(c, L)$ and computed the probability of obtaining a count $B(c, L)$ or a less likely count. For that purpose, we assumed that the resulting distribution of the transform is normal. We computed the raw outlying probability $p_{raw}(c, L)$ of command type $c$ of log $L$ as follows: We assumed that distribution of counts $B(c, L)$ across logs for a command $c$ is normally distributed with probability density function $p$ using the mean $\overline{B(L)}$ and the standard deviation $\sigma(B(L))$. We computed the probability that a random variable $X$ serving as count value is at least (or at most) as large as $B(c, L)$: Given that $B(c, L)$ is larger than the average, ie. $\overline{B(L)} := \sum_{L \in \mathbb{L}} B(c, L)/|\mathbb{L}|$ we computed the probability $p_{raw}(c, L') := p(X > B(c, L'))$ for random variable $X$ and $p_{raw}(c, L) := p(X \leq B(c, L))$ otherwise.

$$\overline{B(L)} := \sum_{L \in \mathbb{L}} B(c, L)/|\mathbb{L}| \tag{2}$$

$$\sigma(B(L)) := \sqrt{\frac{\sum_{L \in \mathbb{L}} (B(c, L) - \overline{B(L)})^2}{|\mathbb{L}|}} \tag{3}$$

$$p_{raw}(c, L) := \begin{cases} p(X > B(c, L)) \text{ if } B(c, L) > \overline{B(L)} \\ p(X \leq B(c, L)) \text{ otherwise} \end{cases} \tag{4}$$

Figure 10 depicts the transformed distribution for frequent command types. The assumption of a normal distribution is not correct for all command types, although the majority of commonly used commands seem to follow a roughly normal distribution. For the rarely used command types, which constitute a significant proportion of all commands (Figure 7), many logs will exhibit a zero count and only a few will have counts larger than zero. Fitting a normal distribution does not assign enough probability mass to the zero value itself or to values that are far from zero. It might be better to model the distribution using a random indicator variable X that states whether the count is zero ($I = 0$) or not ($I = 1$), together with a conditional distribution $p(Y|I = 1)$ that could be normal or more heavily tailed. However, we found that this complexity in modeling is not necessary since rarely used command types have only a little influence on the overall outlier score.

Some deviation from the expected value count of a specific command $B(c, L)$ is supposed to occur. For example, for a standard normal distribution the standard deviation from the mean is 1, so one might not judge a log as being more of an outlier than another log when both are within some limit of the mean. More precisely, we said that all counts within a distance of one standard deviation do not increase the outlier score, ie. the outlier score for a count $X \in [\overline{B(L)} - \sigma(B(L)), \overline{B(L)} + \sigma(B(L))]$ is zero. The cumulative probability density of a value being smaller than the lower limit or larger than the upper limit is given by 0.16. We defined the unweighted outlier score $out(c, L)$ of command $c$ of log $L$ as follows:

$$out(c, L) := \begin{cases} 0 \text{ if } B(c, L) - \overline{B(L)} \in [-\sigma(B(L)), \sigma(B(L))] \\ 1 - p_{raw}(c, L)/0.16 \text{ otherwise} \end{cases} \tag{5}$$

The division by 0.16 serves as normalization so that all values between zero and one are possible.

The distribution of the number of uses of a command type, shown in Figures 7 and 8, shows that it is not uncommon that a command type is used by only a few students. Intuitively, a log that is the only one that uses some command types is likely to be an outlier and so potentially part of plagiarism. We found that honestly created logs often have some unique command types as well. In contrast, in forging a digital product (e.g., rephrasing words, changing layout), standard editing operations seem to suffice, so the corresponding logs do not necessarily contain special commands that are not contained in honestly created logs. However, some command types that are common in honestly created logs are not needed (or are less needed) to disguise a plagiarized work and vice versa. So we added weights for each command that state how much a deviation from the mean is likely to indicate cheating. Generally, the more logs contain the command, the larger its weight. Therefore, a strong deviation in counts from the mean of a command that occurs in all logs could be a strong indicator of an outlier. We computed the weight for a command $c$ as the squared fraction of logs that contain the command. The squaring emphasizes that commands that are not used often should not impact the outlier metric heavily. The computation uses an indicator variable $I$ being one if the command $c$ occurs in log $L$, i.e. $n(c, L) > 0$:

$$I(n(c, L) > 0) := \begin{cases} 1 \text{ if } n(c, L) > 0 \\ 0 \text{ otherwise} \end{cases} \quad (6)$$

$$w(c) := (\frac{\sum_{L \in \mathbb{L}} I(c, L)}{|\mathbb{L}|})^2 \quad (7)$$

The definition yields weights in $[0, 1]$.
The total outlier score for a log $L$ is given by the normalized sum of the contributions of each command:

$$out(L) := \frac{\sum_c w(c) \cdot out(c, L)}{\sum_c w(c)} \quad (8)$$

The larger the score the more likely a log is to be an outlier and, thus, to stem from plagiarism.

## VI. Manual Detection

Automatic detection reveals cheating candidates, but only with varying degree of confidence. Manual inspection of logs can improve clarity for doubtful cases. Our focus is on automatic detection, but we also showcase possibilities for manual inspection that can be classified into more sophisticated analysis of the event logs, questioning the creator, and other techniques.

The histogram-based techniques discussed in Section V are general and require essentially no knowledge about the task being addressed. Detection could be enhanced by performing manual analysis, such as by visually assessing several students' histograms and looking for deviations and similarities. Some commands are typically correlated, such as a large number of "move cursor right" commands that usually also imply as many "move cursor left" commands, "Insert string" commands that correlate with "Delete" commands, and the counts of several commands will change with the amount of navigation and the amount of editing. One can also check to see whether necessary event types are missing, are extremely low or high, or are in the log when they should not be. In short, a more holistic look at the histogram might be beneficial. One might also compare histograms based on a subset of the entire log, which would increase sensitivity and the likelihood of spotting certain behaviors; it would also allow one to compare phases of the creation process that are characterized by different activities, such as when a forged log is a result of appending to a valid log by editing the digital product (e.g., by performing extensive rephrasing of most of the text to mislead conventional plagiarism-detection tools). The distribution of the last part of such a log might be characterized by a large amount of text-replacement events that were performed in a short amount of time. A non-forged log is likely to contain fewer replacements and more navigation or inserts as a student appends to the thesis to finish it or reads through it and fixes typos or changes individual sentences.

In addition to looking at the histograms, one might use additional analysis using logs, such as looking for common subsequences to identify copies and looking for pasted content from external applications. One might also contact the work's creator to inquire about the creation process. For example, the creator should be able to answer questions, such as those related to where we spent the most time on editing and in which order content was created. If keystroke timing is available, one might also conduct simple tests like keystroke pattern analysis to, for example, determine whether the time between keystrokes is similar during a supervised test to the time between keystrokes that is recorded in the submitted log [10]. Another option is to consider existing techniques for verification, such as computing the similarity in the final source code using the MOSS tool [29] or, for a text document, checking the consistency of multiple works by the same author [19]. Finally, one might also use techniques that are not related to creation logs to determine whether a creator is cheating. We refer to the related work (Section II) for such techniques.

## VII. Implementation And Evaluation

### A. Collected Data

We used the Eclipse IDE with the Fluorite logger plug-in [36] to obtain logs from more than sixty students for three programming assignments. We conducted the same analysis for all three assignments, but all gave comparable results, so we discuss only the first assignment in detail. Given a skeleton of multiple files, the students were supposed to write roughly 100 lines of code.

Figure 7 shows the frequencies of certain types of commands. A few commands that were used frequently were related mainly to editing and navigation. Figure 8, which shows how many students used a certain command, reveals that about 10 percent of all commands were used by all of the students, and about 50 percent were used by no more than three students. Command use might also be unintentional, such as when a student chose a menu item by accident. The use of a particular command type varied widely from student to student. Figure 9, which shows the box plot of the fifteen most frequent event types, reveals that the student at the 25th percentile used a particular command roughly a factor 2-15 less than the student at the 75th percentile did. Some of the submitted logs were short because they came from incomplete assignments. Some of the logs also contained work that was not part of the assignment. We did not filter these logs, although they are likely to have a negative effect on the outlier-detection result.

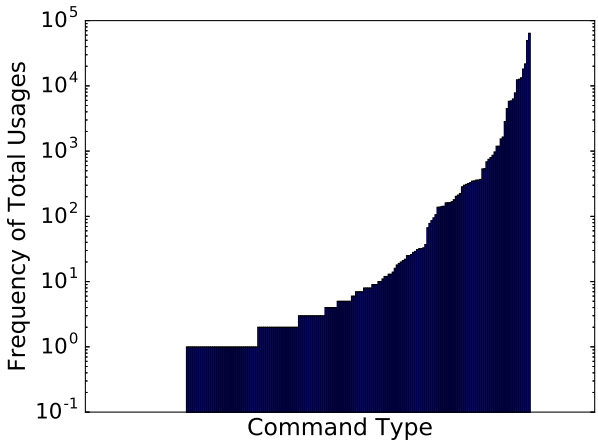

Fig. 7. Distribution of use of command types by all students

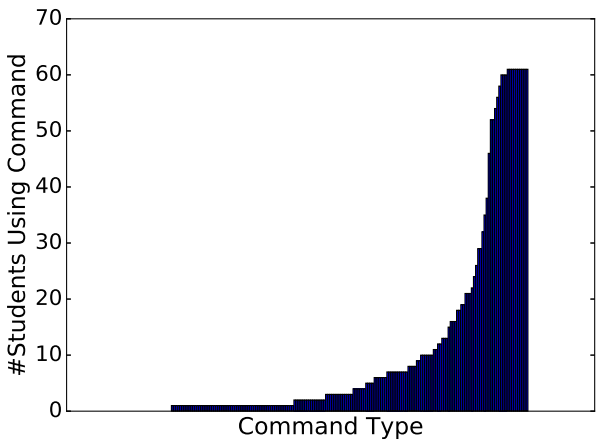

Fig. 8. Distribution of the number of students who use a command type

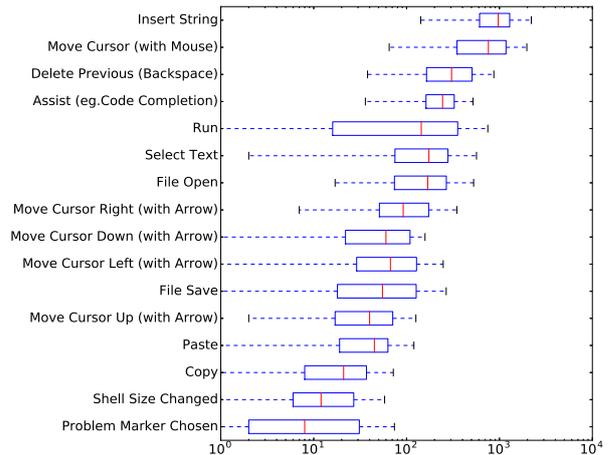

Fig. 9. Box plot of counts of 16 most frequent event types.

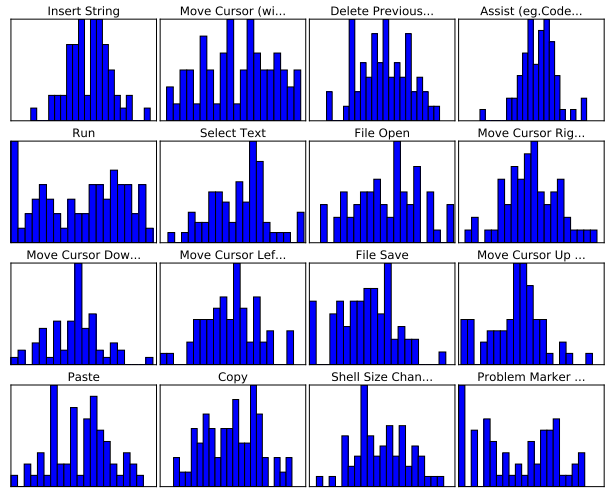

Fig. 10. Distribution of the 16 most frequent event types after Box-Cox transformation.

### B. "Plagiarism" Data

Dataset 1: All of the logs collected from students were used to conduct a semi-automatic detection process that identified potential cheaters automatically and then checked the logs manually to determine whether they were, indeed, cheaters.

Dataset 2: Dataset two is an extension of Dataset 1, as it adds synthetic data created through modification from the original logs without using the IDE. The dataset should help in addressing two questions: What is the minimum required change so that a log is not detected as being a modified copy of another? How much does a normal log have to be modified in order to be considered an outlier? To address these questions, each log was modified in two ways:

- *Event type* change: We varied the number of event types by removing from the logs a percentage of all of some types of events (chosen uniformly at random). This modification addresses the case in which a cheater is not using all kinds of events because he copied and pasted the final outcome.

- *Event frequency* distribution change: We altered the distribution of events by increasing or decreasing the count of events. Given a maximal change factor $k$, we changed the logs' counts of each command type by choosing a random number $r \in [1, k]$ and increasing the count in half the logs by a factor $r$ and decreasing the count in the other half of the logs by a factor of $1/r$. This approach is in response to the chance that a cheater will use certain commands more often or less often than most of his peers (e.g., more copy-and-paste, fewer undos, editing, and debugging). This approach covers the most appealing scenario for plagiarism: Taking an honestly created log and appending to it by editing the digital product.

For each log and each change factor $k$ we created ten modified versions.

Dataset 3: This dataset extends Dataset 1 by using the IDE to create a plagiarized solution to an assignment. We considered several strategies for obtaining a "fake" log from scratch–that is, without using an existing log–that require increasing effort from the cheater. Our strategies are similar to [12], which focuses on Wikipedia articles.

- Copy-and-paste: Copying the source code (e.g., from the Internet) without modification or any additional activity,

such as running the code.

- **Small Refactoring:** Copying the source code and renaming a few variables.
- **Medium Refactoring:** Copying the source code, renaming several variables, changing some comments, and running the code. The amount of work performed is roughly a factor of ten less than the work shown in the average log.
- **Large Refactoring:** Copying the source code, renaming most variables, changing many of the comments, reordering statements, making minor code changes, and running the code to test. The size of the log is slightly larger than the average log size.

We created three logs manually for each strategy, for a total of twelve logs. Then we created 100 variations of each of the twelve manually created logs by changing the event frequency using a maximal change factor of three, as described for the event-frequency alterations of Dataset 2.

### C. Setup and Parameters

Using a PC (2.7 GHz CPU, 8 GB RAM), we ran our histogram-based detection on our collected data (all students, all three assignments). To compute the similarity between two logs, we chose $n_{sam.} = 100$ subsets of commands of size $s_{sam.}$ of 2, 4, 8, and 16.

**Experiment using Dataset 1:** The objectives of this experiment were to identify inadequate logs (e.g., incompletely submitted logs) and their impact. We identified logs that appeared to be outliers and those that appeared to be very similar for reasons that range from cheating to students' erroneous handling of logs (e.g., not submitting all log data or submitting an excessive amount of log data that included work beyond the assignments).

In looking for cheaters in Dataset 1, we selected the pair(s) of students with maximal similarity in the correlation measure (Equation 1) and those with the largest outlier score (Equation 8) as cheaters. We manually inspected the ten most similar pairs of logs and the ten most outlying logs.

**Experiment using Dataset 2:** Dataset 2 builds on the assumption that there are no cheaters in Dataset 1. The goal of the experiment was to assess the sensitivity of the detection method for the level of a log's deviation from another log or the set of all logs. For each log, we created a modified log and then ran the detection algorithm against all original logs and the modified log. Thus, for each modified log we obtained a correlation score (Equation 1) and an outlier score (Equation 8). An instructor might have only little time to check for plagiarism that is used to investigate a few high-risk candidates for cheating. Therefore, we said that a cheater was detected if his or her log was among the five most similar logs (for a cheater who copies and modifies a log) or among the five most dissimilar logs (for a cheater who creates a log from scratch). We computed the percentage of faked logs that were detected and assessed the detection capability for both kinds of synthetic data (i.e., removing some event types completely and gradually changing the distribution of events).

**Experiment using Dataset 3:** The goal of this experiment was to determine whether the detection method can detect logs that are created from scratch. As in Experiment 2, we added each of the created (cheating) logs to the entire set of logs and tested whether its outlier score or correlation score ranked among the five highest.

### D. Results

The entire computation took less than one hour.

**Parameters:** The results of all assessments were best for $s_{sam} = 2$. This is no surprise since, if a cheater does not alter the frequency of all of the command types, it is likely that the correlation is maximal (i.e., 1) because any two commands will have the same counts. If we pick more commands than two, different counts become more likely, yielding a smaller correlation.

**Results Experiment 1:** The distribution of the distances of original logs is shown in Figures 11 and 12. Figure 11 shows the correlation scores used to identify copies of an assignment (i.e., $cor(L, L')$ as defined in Equation 1), while outlier scores (i.e., $out(L, L')$ as in Equation 8), are shown in Figure 12. Because of variations in individual students' behavior, gaps are expected, but the larger the gap, the more suspicious the work. For the original dataset, we judged that the ten most similar pairs of logs were created by different students and found the ten most dissimilar logs to be logs that came from students who were solving the assignment honestly. Therefore, we concluded that there are likely no cheaters in this dataset. Some logs did not provide a completed assignment (i.e., students gave up and submitted incomplete assignments), so we did not check such logs for plagiarism. We employed several checks as described in Section VI and looked at each log's distribution of commands with respect to others' distribution. For example, we created a figure for each log $L$ (Figure 9) that contained more event types and highlighted the counts for log $L$. We also did this for partial logs but found no suspicious patterns.

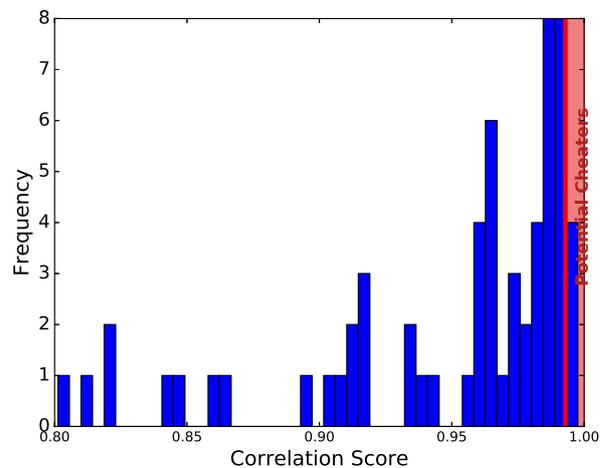

Fig. 11. Histogram of pair-wise histogram distances $cor(L, L')$. The red line shows the threshold for being a potential cheater.

**Results Experiment 2:** When we added synthetically created logs, we saw an expected trade-off in both datasets: The more distorted a student's log, the less likely it was detected as a copy and the more likely it was detected as an outlier.

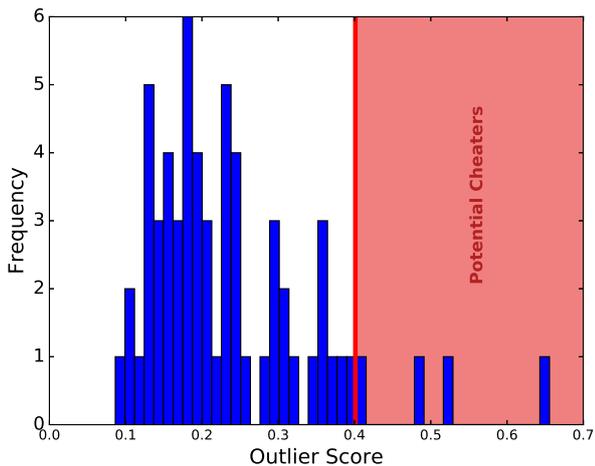

Fig. 12. Histogram of pair-wise histogram distances $out(L, L')$. The red line shows the threshold for being a potential cheater.

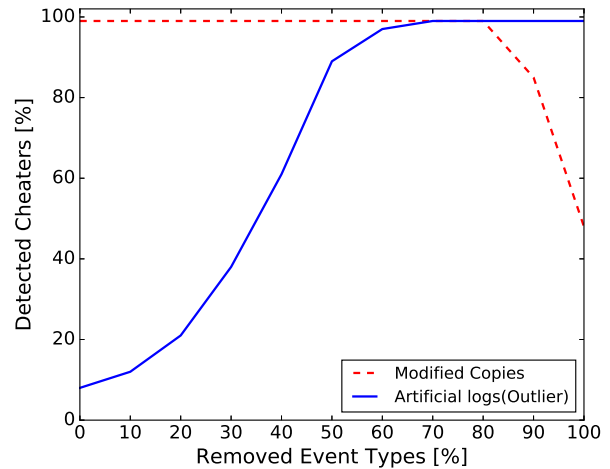

Fig. 13. Fraction of faked logs detected for event-type change.

Detection of modified copies: When we removed 90 percent of all event types (Figure 13) or changed event counts by a factor of up to 10 (Figure 14), we still detected about 90 percent of copied and modified logs as plagiarism. The high detection rate occurred because it suffices to find a pair of command types that were not modified. This chance is relatively high even if many command types are altered: Say we removed 90 percent of command types for a log and left the counts of the other 10 percent unchanged. For computation of similarity we chose one command type from the original log and the modified log with a count larger than zero. The one from the modified log occurs in the original, but the one from the original log occurs in the modified log with only 10 percent probability, so there is a 10 percent probability that the counts are identical (i.e., perfect correlation) for one subset. If we modify counts by a random factor, there is still a reasonable chance that a few command types are altered by a small amount, which results in a high detection rate.

Detection of artificially created copies: Outlier detection requires relatively large changes of frequencies (see Figure 14). When modifying a log created by an honest student, we must change it beyond the variance across all logs first before it becomes an outlier. Given the fact that logs vary strongly, partially due to incomplete or incorrect logs (see Section VII-A), the change required is also significant, ie. when changing command execution counts by a factor of 5, we detect about 40% of logs as outliers. For removing event types similar reasoning applies, but detection seems to work better because removing a few frequent event types that occur in most logs has a strong impact on the outlier score. We classify about 90% of logs as outliers, if they lack 50% of events, while being otherwise identical to one of the logs.

Results Experiment 3: Detection worked well in all cases (Figure 15) since the logs of plagiarized work were likely to lack entirely or to a large extent some frequently used commands (e.g., commands for opening files, edits, navigation, saving, executing the source code).

Although our results are encouraging, our work has some limitations, as discussed in the next section.

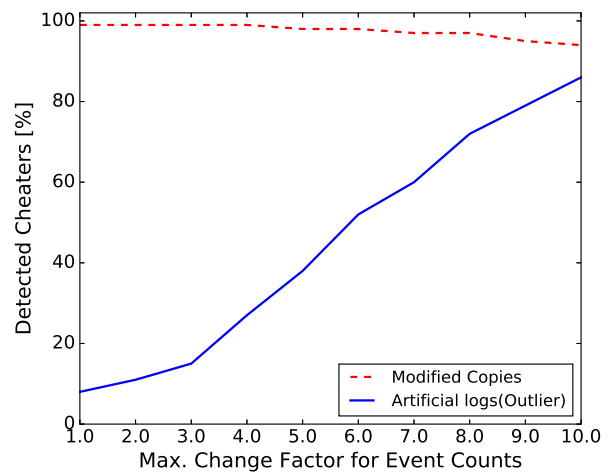

Fig. 14. Fraction of faked logs detected for event frequency distribution change.

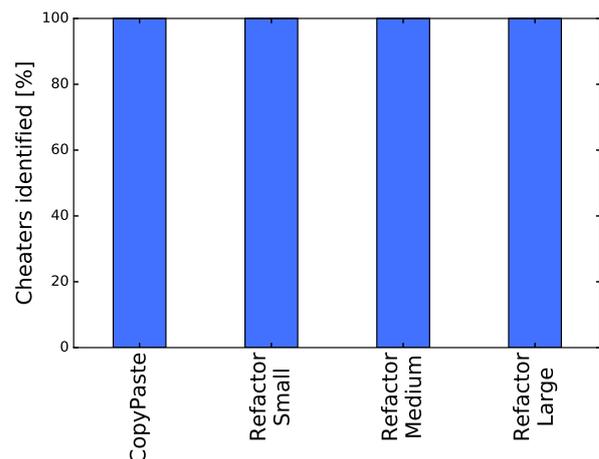

Fig. 15. Detection of cheaters with variations of logs created using various approaches.

## VIII. DISCUSSION AND FUTURE WORK

We mention the advantages and disadvantages of using the creation process to detect plagiarism, as well as suggestions for future work that targets improving automatic detection and facilitating manual inspection of logs.

### A. Strengths and Weaknesses

We identified several reasons why plagiarism detection based on logs is attractive:

- Amount of information: Logs typically contain much more information than the final product does, since logs usually implicitly include the final work as well as all actions and changes during the creation process (e.g., editing, saving, and printing). Since logs have more information, they lend themselves to accurate methods of detection. For that reason, forging a log requires more effort than does forging the final product. Our approach forces cheaters not only to modify copied content, as is required for traditional methods of hiding plagiarism, but it also demands that they understand and mimic the creation process.

- Complexity of log forging: Creating a forged log often requires in depth understanding of the log-creation software. Log entries often depend on internal processes that are hard to predict, such as the exact time of automatic software updates and the number of lines that result from inputting a text in a text editor.

- Novel detection techniques: A whole set of new techniques from data mining that are based on outlier detection or nearest-neighbor approaches can support the detection of plagiarism based on logs. Our technique is data-driven with direct comparisons among a set of logs, rather than relying on expert input or dedicated rules. This approach makes our technique flexible so it is easily usable for logs that come from different programs and different tasks from different domains that range from natural sciences to economics to sociology. Our approach may also detect plagiarized work that other techniques that focus on the final product fail to identify.

- Structure of logs: In contrast to the digital outcome, which might appear as "unstructured" data, such as text from a data science perspective, logs appear as structured data, facilitating their analysis.

Other potential advantages are that logs can also be used for other purposes that might increase learning, including giving students automated feedback that allows them reflect on their behavior and capabilities and to see recommended functions used by fellow students. Knowing that logs are created might also encourage students to work in a more focused way.

Log-based detection also comes with certain weaknesses compared to conventional plagiarism-detection techniques. If logs are similar, the odds of a false positive are still rather small, but the risk of false positives are higher for students who behave differently from the majority of students. For example, if a student happens to require much less rework than other students do, he risks being classified as potential cheater. The same holds for students who perform a large amount of editing to improve the quality of their work. Therefore, our detection technique often requires manual interaction, such as manual examination of logs that seem to be forged. Although this issue might also hold for conventional techniques, the concept of logs might seem harder for an instructor to grasp and assess, at least in an initial phase.

In our classes, we require that students collect and submit logs by installing a plugin that records events. Installation of the plugin takes less than five minutes and submission of a log less than two minutes. Students must disable the logging when they work on private projects. Both limitations could be reduced or eliminated if, for example, one installs a private version of the software (without the logging plugin) and university version (with the logging plugin).

Students may also have to use a particular software that supports logging for creating their work. In practice, it is not uncommon for students to be required to use certain applications, such as templates for theses that are available only for a particular text editor or a course that requires the student to submit project files from a particular editor that was also used in tutorials. In principle, one might use a HCI recorder or support multiple applications, but doing so increases the effort required from the lecturer.

In addition, although many applications support logging, the details of such logs differ significantly, which might limit the ability to replay a log to create the digital outcome. For example, whereas the macro recorder in Microsoft Excel logs mouse clicks in a spreadsheet, the recorder in Microsoft Word logs only keystrokes and clicks on buttons. In the Excel case, the (final) spreadsheet can be created by replaying the log, but in the Word case, while one might check the log for semantic correctness by replaying it, it might not be possible to create the final document by replaying the log. Still, the plagiarism-detection techniques presented here can be employed in Word by using the in-built recorder to create logs, although a cheater might forge logs more easily if no additional means are undertaken to check whether the submitted log corresponds to the submitted digital outcome.

### B. Possible Improvements

There are many techniques other than using histograms, such as identifying a specific cheating behavior based on rules like copy-and-paste and paraphrasing. Typically, copying of text and images occurs outside the creation software (e.g., in a web-browser), and the pasting occurs in the creation software. A large amount of pasting in terms of both frequency and quantity might be an indicator of cheating, and inserting a lot of text in a short amount of time with relatively little editing (e.g., delete events) might be an indicator of paraphrasing. Fingerprinting of keystrokes [10] is another way to identify forged logs. More generally, one could improve the detection mechanism to incorporate the timing of events. One might also look at short sequences of events by, for example, choosing a sequence in a log that seems unique (any sufficiently long sequence is unique) and searching for this sequence in other logs to determine whether the log was copied.

Though all techniques for detecting plagiarism apply to any kind of digital outcome, we evaluated our technique on programming assignments conducted in a complex IDE. The process and tools used for programming are arguably more complex than are those used to create text documents for assignments that involve merely expressing ideas and summarizing work, without actual implementation. For example, activities like testing and debugging do not occur in ordinary text-processing, so the set of commands that are used when creating the digital outcome is reduced. Therefore, forging a log created in a simple tool with a less diverse command set might be easier, requiring more advanced detection techniques. Histograms that use short sequences of events rather than single commands are a plausible option.

We focused on assignments solved by many students, rather than theses, where students work on a variety of topics, as logs of assignments might be more homogeneous than logs of thesis. Homogeneous logs give a cheater fewer options to forge a log without being detected since homogeneous logs are characterized by little variance in usage statistics. Theses also adhere to a common structure and also have specific creation processes, and they are significantly longer, resulting in longer log files. Here, an approach that examines sequences of events related to a section of a thesis, such as the literature review or the introduction, rather than a single large log, might improve detection accuracy.

Whereas simple cheating attempts that involve copy-and paste and some textual changes are easily detected, we do not currently protect well against a student who manually enters text and thoroughly simulates the creation process by pasting copied text and then conducting artificial editing. If the amount and way of editing are similar to those other logs, a cheater might well escape detection. However, without automatic generation of logs, the manual work in which a student must invest significantly increases compared to the work required to alter the final outcome. The amount of work (measured in interactions with the tool) seems to be similar to that of peers who work honestly.

To improve detection and support the inspection of logs, we envision a tool that visualizes the creation process in an intuitive manner. Such a tool would allow a skilled person, such as a thesis supervisor or a teaching assistant, to identify easily whether a semantically correct log is likely to have been forged. For instance, finding that a difficult part of the work was done fairly quickly, while other, simpler parts required a lot of time could make the work suspicious. A tool that visualizes the creation process in an intuitive manner might help to avoid false positives by gathering evidence that supports whether people who are suspected of plagiarism are, indeed, guilty.

## IX. CONCLUSIONS

Cheating has always been an issue both inside and outside education. We contribute to remedying this problem by proposing a process for automatic identification of likely cheaters. Our novel approach requires using readily available programs (with appropriate plug-ins) to collect logs, rather than just the final digital result. Ideally, our detection technique will be combined with other techniques that analyze the final "digital" outcome, which would make cheating time-consuming and impractical. Of course, automatic tools that can circumvent log-based detection are certain to be developed, so catching offenders will remain a cat-and-mouse game. Even so, the game must be played to counteract incentives for cheating and to ensure that cheaters are not among the graduates of universities who obtain powerful positions, where misconduct can harm large parts of society. We see this work as a step in this direction.

**Acknowledgments:** We would like to thank YoungSeek Yoon (author of Fluorite[36]), Shen Gao and Perekrestenko Dmytro for valuable discussions.

## REFERENCES

[1] A. Ahadi, R. Lister, H. Haapala, and A. Vihavainen. Exploring machine learning methods to automatically identify students in need of assistance. In *Proc. of Conf. on Int. Computing Education Research*, pages 121–130, 2015.

[2] M. Alsallal, R. Iqbal, S. Amin, and A. James. Intrinsic plagiarism detection using latent semantic indexing and stylometry. In *Int. Conf. on Developments in eSystems Engineering (DeSE)*, pages 145–150, 2013.

[3] S. M. Alzahrani, N. Salim, and A. Abraham. Understanding plagiarism linguistic patterns, textual features, and detection methods. *IEEE Transactions on Systems, Man, and Cybernetics, Part C (Applications and Reviews)*, 42(2):133–149, 2012.

[4] V. Anjali, T. Swapna, and B. Jayaraman. Plagiarism detection for java programs without source codes. *Procedia Computer Science*, 46:749–758, 2015.

[5] A. Bin-Habtoor and M. Zaher. A survey on plagiarism detection systems. *Int. Journal of Computer Theory and Engineering*, 4(2):185–188, 2012.

[6] P. Blikstein, M. Worsley, C. Piech, M. Sahami, S. Cooper, and D. Koller. Programming pluralism: Using learning analytics to detect patterns in the learning of computer progr. *Journal of the Learning Sciences*, 23(4):561–599, 2014.

[7] T. Bliss. Statistical methods to detect cheating on tests: A review of the literature. *National Conference of Bar Examiner (NCBE)*, 2012.

[8] A. Caliskan-Islam, R. Harang, A. Liu, A. Narayanan, C. Voss, F. Yamaguchi, and R. Greenstadt. De-anonymizing programmers via code stylometry. In *USENIX Security Symposium*, pages 255–270, 2015.

[9] D.-K. Chae, J. Ha, S.-W. Kim, B. Kang, and E. G. Im. Software plagiarism detection: a graph-based approach. In *Proc. of Conf. on Information & knowledge management (CIKM)*, pages 1577–1580, 2013.

[10] T.-Y. Chang, C.-J. Tsai, Y.-J. Yang, and P.-C. Cheng. User authentication using rhythm click characteristics for non-keyboard devices. In *Proc. of Conf. on Asia Agriculture and Animal (IPCBEE)*, volume 13, pages 167–171, 2011.

[11] D. Chuda, P. Navrat, B. Kovacova, and P. Humay. The issue of (software) plagiarism: A student view. *IEEE Transactions on Education*, 55(1):22–28, 2012.

[12] P. Clough and M. Stevenson. Developing a corpus of plagiarised short answers. *Language Resources and Evaluation*, 45(1):5–24, 2011.

[13] G. Cosma and M. Joy. An approach to source-code plagiarism detection and investigation using latent semantic analysis. *IEEE Transactions on Computers*, 61(3):379–394, 2012.

[14] K. Damevski, D. Shepherd, J. Schneider, and L. Pollock. Mining sequences of developer interactions in visual studio for usage smells. *IEEE Transactions on Software Engineering*, 99:1–14, 2016.

[15] B. Gipp and N. Meuschke. Citation pattern matching algorithms for citation-based plagiarism detection. In *Proc. of symposium on Document engineering*, pages 249–258, 2011.

[16] M. Jiffriya, M. A. Jahan, H. Gamaarachchi, and R. G. Ragel. Accelerating text-based plagiarism detection using gpus. In *Int. Conf. on Industrial and Information Systems (ICIIS)*, pages 395–400, 2015.

[17] M. Joy, G. Cosma, J. Y.-K. Yau, and J. Sinclair. Source code plagiarisma student perspective. *IEEE Transactions on Education*, 54(1):125–132, 2011.

[18] M. Kersten and G. C. Murphy. Using task context to improve programmer productivity. In *Proc. of symposium on Foundations of software engineering*, pages 1–11, 2006.

[19] J. Li, R. Zheng, and H. Chen. From fingerprint to writeprint. *Communications of the ACM*, 49(4):76–82, 2006.

[20] R. Lukashenko, V. Graudina, and J. Grundspenkis. Computer-based plagiarism detection methods and tools: an overview. In *Proc. of the int. conference on Computer systems and technologies*, pages 40–, 2007.

[21] H. A. Maurer, F. Kappe, and B. Zaka. Plagiarism-a survey. *Journal of Universal Computer Science*, 12(8):1050–1084, 2006.

[22] A. M. Memon. Gui testing: Pitfalls and process. *IEEE Computer*, 35(8):87–88, 2002.

[23] N. Meuschke and B. Gipp. State-of-the-art in detecting academic plagiarism. *Int. Journal for Educational Integrity*, 9(1), 2013.

[24] N. Meuschke, B. Gipp, C. Breitinger, and U. Berkeley. Citeplag: A citation-based plagiarism detection system prototype. In *Proc. of Int. Plagiarism Conference*, 2012.

[25] M. Novak. Review of source-code plagiarism detection in academia. In *Conv. on Inf. and Com. Tech., Electronics and Microel.*, pages 796–801, 2016.

[26] M. Potthast, B. Stein, A. Barrón-Cedeño, and P. Rosso. An evaluation framework for plagiarism detection. In *Proceedings of the 23rd international conference on computational linguistics: Posters*, pages 997–1005. Association for Computational Linguistics, 2010.

[27] J. A. Reither. Writing and knowing: Toward redefining the writing process. *College English*, 47(6):620–628, 1985.

[28] F. Rosales, A. García, S. Rodríguez, J. L. Pedraza, R. Méndez, and M. M. Nieto. Detection of plagiarism in programming assignments. *IEEE Transactions on Education*, 51(2):174–183, 2008.

[29] S. Schleimer, D. S. Wilkerson, and A. Aiken. Winnowing: local algorithms for document fingerprinting. In *Proceedings of the 2003 ACM SIGMOD international conference on Management of data*, pages 76–85. ACM, 2003.

[30] M. Schonlau, W. DuMouchel, W.-H. Ju, A. F. Karr, M. Theus, and Y. Vardi. Computer intrusion: Detecting masquerades. *Journal of Statistical science*, pages 58–74, 2001.

[31] J. Shah, A. Shah, and R. Pietrobon. Scientific writing of novice researchers: what difficulties and encouragements do they encounter? *Academic Medicine*, 84(4):511–516, 2009.

[32] C. Simmons. Codeskimmer: a novel visualization tool for capturing, replaying, and understanding fine-grained change in software. *http://hdl.handle.net/2142/44125*, 2013.

[33] W. Snipes, A. R. Nair, and E. Murphy-Hill. Experiences gamifying developer adoption of practices and tools. In *Companion Proc. of the 36th International Conference on Software Engineering*, pages 105–114, 2014.

[34] A. Vihavainen, J. Helminen, and P. Ihantola. How novices tackle their first lines of code in an ide: analysis of programming session traces. In *Proc. of the Int. Conf. on Computing Education Research*, pages 109–116, 2014.

[35] M. J. Wise. Yap3: improved detection of similarities in computer program and other texts. In *ACM SIGCSE Bulletin*, volume 28, pages 130–134, 1996.

[36] Y. Yoon and B. A. Myers. Capturing and analyzing low-level events from the code editor. In *Proc. of the 3rd SIGPLAN workshop on Evaluation and usability of programming languages and tools*, pages 25–30, 2011.

[37] F. Zhang, Y.-C. Jhi, D. Wu, P. Liu, and S. Zhu. A first step towards algorithm plagiarism detection. In *Proc. of the Int. Symposium on Software Testing and Analysis*, pages 111–121, 2012.

[38] M. Zurini. Stylometry metrics selection for creating a model for evaluating the writing style of authors according to their cultural orientation. *Informatica Economica*, 19(3):107, 2015.